\documentclass[preprint,aps]{revtex4}

\usepackage{graphicx}
\usepackage{dcolumn}
\usepackage{bm}

\begin{document} 
\newcommand{\vk}{{\vec k}} 
\newcommand{\vK}{{\vec K}}  
\newcommand{\vb}{{\vec b}}  
\newcommand{{\vp}}{{\vecp}}  
\newcommand{{\vq}}{{\vec q}}  
\newcommand{\vQ}{{\vec Q}} 
\newcommand{\vx}{{\vec x}} 
\newcommand{\vh}{{\hat{v}}} 
\newcommand{\tr}{{{\rm Tr}}}  
\newcommand{\beq}{\begin{equation}} 
\newcommand{\eeq}[1]{\label{#1} \end{equation}}  
\newcommand{\half}{{\textstyle\frac{1}{2}}}  
\newcommand{\gton}{\stackrel{>}{\sim}} 
\newcommand{\lton}{\mathrel{\lower.9ex 
  \hbox{$\stackrel{\displaystyle <}{\sim}$}}}  
\newcommand{\ee}{\end{equation}} 
\newcommand{\ben}{\begin{enumerate}}  
\newcommand{\een}{\end{enumerate}} 
\newcommand{\bit}{\begin{itemize}}  
\newcommand{\eit}{\end{itemize}} 
\newcommand{\bc}{\begin{center}}  
\newcommand{\ec}{\end{center}} 
\newcommand{\bea}{\begin{eqnarray}}  
\newcommand{\eea}{\end{eqnarray}} 
\newcommand{\beqar}{\begin{eqnarray}}  
\newcommand{\eeqar}[1]{\label{#1} 
\end{eqnarray}}  
 
\title{Does the Charm Flow at RHIC?}
 
\author{S. Batsouli, S. Kelly, M. Gyulassy}
\affiliation{Columbia University \\ 538 West 120th Street \\ New York, NY 10027}

\author{J.L. Nagle}
\affiliation{University of Colorado at Boulder \\ 2000 Colorado Avenue \\ Boulder, CO 80309}

\begin{abstract} 
Recent PHENIX $Au+Au\rightarrow e^- +X$ data~\cite{PHENIXe} 
from open charm decay are shown to be consistent 
with two extreme opposite 
dynamical scenarios of ultra-relativistic nuclear reactions.
Perturbative QCD {\em without} final state interactions
was previously shown to be consistent with the data. 
However, we show that the data are  also 
consistent with zero mean free path
hydrodynamics characterized by a common transverse flow velocity
field. The surprising coincidence of both $D$ and $B$
hydrodynamic flow spectra 
with pQCD up to $p_T \approx 3$ and $5$ GeV, respectively,
suggests that heavy quarks {\em may} be produced essentially at
rest in the rapidly expanding gluon plasma. Possible implications and
further tests of collective heavy quark dynamics are discussed.
\vspace{.2cm} 
\end{abstract} 
 
\maketitle

 
\section{Introduction} 

Heavy quark production in nuclear collisions ($A+A\rightarrow c(b) +X$)
is conventionally calculated via 
pQCD~\cite{Combridge:1978kx,Gavai:1994gb,Sjostrand:2000wi,Wang:1991ht}.
Final state elastic scattering effects either at the partonic ~\cite{Thoma:1990fm} or the hadronic rescattering level~\cite{Zhang:2002ta,Lin:2000jp}
are not expected to be large 
distortions of the initial spectra 
because the cross sections involved are small.
On the other hand, energy loss in dense matter via induced gluon radiation 
could possibly lead to a
large suppression of the high $p_T$
distribution relative to pQCD predictions~\cite{Shuryak:1996gc,Mustafa:1997pm,Lin:1998bd}.  
It was this suppression that was thought to be essential to prevent 
the open charm ``background'' from swamping
dilepton signatures of the sought after thermal plasma.
However, more recent considerations suggest that
the ``dead cone'' effect for heavy quarks~\cite{Dokshitzer:2001zm}
may inhibit induced radiative energy loss. More quantitative calculations 
appear to support this assertion~\cite{magda}.
Therefore, it is not unreasonable to expect 
that  heavy quark (and open charm hadron) production
could be well approximated by conventional
factorized pQCD even in central $Au+Au$ collisions at RHIC. 
This would indicate that the produced medium is essentially 
transparent to heavy quarks, or equivalently, the 
heavy quark and subsequent D or B have 
mean free paths at least as large as nuclear dimensions.
In this case, the charm and bottom quark yields are predicted  to
scale with atomic number $A$ and impact parameter $b$, simply according to the 
Glauber nuclear overlap density $T_{AB}(b)$. 
The predicted  distribution of heavy quarks, $Q=c,b$, is then
\begin{equation}
\frac{dN^{AB(b)}_Q}{dyd^2pt}  = T_{AB}(b) \frac{d\sigma^{PQCD}_Q}{dyd^2pt}
\;\; .\end{equation}
For central 10\% $Au+Au$ collisions $T_{AuAu}(cent) = 22.6~mb^{-1}$. Nuclear
geometry therefore  amplifies
the $p + p$ rate by the number of binary collisions $N_{binary} \approx 905$.
See ref. \cite{Gavai:1994gb} for an extensive survey of pQCD heavy quark production.

Experimentally, recent PHENIX data~\cite{PHENIX_e} on ``prompt'' 
single electron
production in central and ``minimum bias'' $Au + Au$ collisions at $\sqrt{s}=130$
AGeV have tested the above pQCD predictions.   Utilizing
PYTHIA~\cite{Sjostrand:2000wi} to take into account the $c\rightarrow
D, D^*$ fragmentation and subsequent open charm decays, the data were
found to be in good agreement with the pQCD predictions within the
errors quoted, revealing no indication of a medium effect.
Furthermore the observed binary collision scaling of the yield with centrality
shows no hint (within relatively large errors) of possible suppression due to gluon shadowing or
saturation effects.  It should be noted that nuclear modifications to the
parton distribution functions may only lead to a small effect on
the charm cross section because the PHENIX acceptance spans 
both the conventional shadowing and anti-shadowing $x\sim 0.05-0.15$ range.

In this letter we point out however that the
single electron spectra in the observed $p_T< 3$ GeV range may hide
much more extreme dynamics. We show 
that the same observed $e^-$ spectra is 
reproduced by a thermal hydrodynamic
model that is consistent with the lighter hadron transverse momentum distributions.
In the  $p_T < 2$ GeV range, 
striking collective flow signatures have been observed at RHIC
for  $\pi, K, p$~\cite{v2phenix,v2star,ptdata}. Hydrodynamic
models of course assume that (at least for light quarks and gluons) 
the opacity of the produced quark-gluon plasma is high enough 
that local equilibrium is achieved early in the collision 
and maintained through hadronization. 
This extreme assumption, first proposed by  Landau and Feinberg,
has until recently  consistently
over-predicted collective flow effects in hadronic interactions.

The collective flow velocity field in hydrodynamics 
is predicted to lead to strong mass dependent distortions of the
transverse momentum spectra~\cite{Stocker:jr,Shuryak:ds,VanHove:1982vk} 
relative to pQCD predictions in Eqn. 1. In pQCD the spectral distribution
 in $A + B$ collisions is identical to that in $p + p$ up to the $T_{AB}$ geometrical
scale factor.  Recent detailed hydrodynamic calculations
of collective flow patterns expected in $Au+Au$ at $\sqrt{s}=130,200$ AGeV 
using realistic QCD equations of state
found  remarkable good agreement with the radial flow and 
azimuthal asymmetries observed in the $\pi,K,p$ 
spectra~\cite{Huovinen:2001cy,Teaney:2001av,Huovinen:2002rn}. 
This is in sharp contrast to lower energy data (SPS, AGS, SIS, Bevalac)
where one fluid hydrodynamics always over-predicted collective flow
and large dissipative non-equilibrium corrections to its predictions 
were necessary to reconcile theory with data. We note that
phenomenological fits~\cite{Schnedermann:1993ws,Peitzmann:2002nt}
with adjustable parameters could reproduce SPS data,
but these fits required fine tuning of initial conditions 
and are not consistent with more realistic 
one-fluid hydrodynamics results~\cite{Kolb:2000fh}.

At RHIC the produced density and equilibration rates 
could be high enough
that at least the light quarks and gluons may achieve 
local equilibrium and subsequent hydrodynamic collective flow.
However, that local equilibrium could be achieved poses an entirely non-trivial
problem for transport theory. Recent studies
~\cite{Molnar:2001ux,Lin:2002gc,Bass:2002fh,Dumitru:2000up}
based on parton cascade dynamics indicate that the opacity  
needed to achieve local equilibrium must be at least 
an order of magnitude higher
than pQCD estimates.  
We note that non-linear Yang-Mills 
dynamics~\cite{Teaney:2002kn} is too weak to account for 
the strong observed azimuthal collectivity observed out to several GeV.
In contrast, the data are easily described {\em if}
 local equilibrium is simply {\em assumed}! Jet tomographic analysis of the
high $p_T$ quenching pattern also suggests that 
the matter produced is indeed highly opaque~\cite{Gyulassy:2001nm}.

The question then naturally arises whether in spite of the strong
theoretical prejudice against heavy quark 
local equilibration in nuclear collisions, 
could the mechanisms that appear to effect such an equilibration in the light partons also
coerce the heavy quarks ``to go with the flow''~\cite{Bugaev:2002fd}?
We test this hypothesis by computing the transverse momentum spectra of open 
charm and bottom hadrons using the same transverse boosted Bjorken 
model that fits the light hadron $p_T$ spectra up to 2 GeV.
The numerical results of realistic Bjorken boost invariant hydrodynamics
at RHIC can be approximated via~\cite{Schnedermann:1993ws,Kolb:2000fh}
%
%
\begin{equation}
\frac{dN_H}{dyd^2p_\perp}~=~ \frac{dN_H}{dy} \frac{m_\perp}{Z}\int_0^R r dr 
I_0 \left( \frac{p_\perp \sinh \rho_\perp(r)}{T_{fo}} \right) 
K_1 \left( \frac{m_\perp \cosh \rho_\perp(r)}{T_{fo}} \right) 
\end{equation}
where $T_{fo}$ is the freeze-out temperature and $Z$ normalizes the transverse momentum integral.
The radial Doppler boost rapidity is taken as  
\begin{equation}
\rho_\perp(r) = \tanh ^{-1}(\beta_{T}(r))
\end{equation}
and 
\begin{equation}
\beta_{T}(r) = \beta_{max} \left( \frac{r}{R} \right)
\end{equation}
which assumes a linear boost profile.



\begin{figure}
\includegraphics[height=.5\textheight]{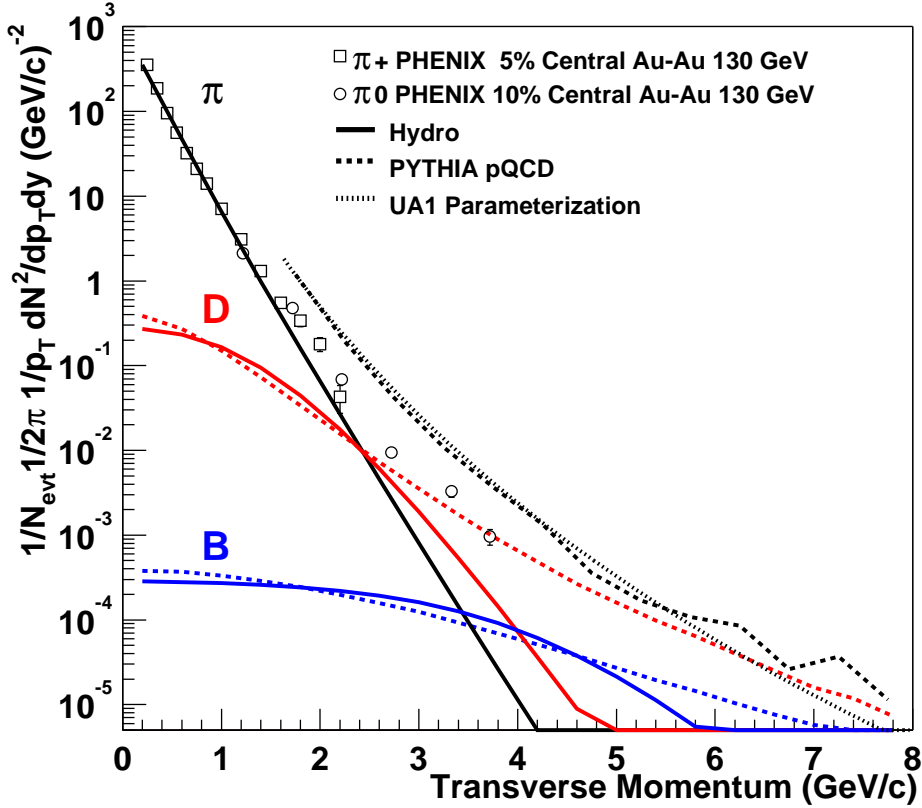}
\caption{For 10\% central $Au + Au$ collisions at $\sqrt{s} = 130$ AGeV we show the neutral pion $\pi^{0}$, $D$ meson, 
and $B$ meson transverse momentum distributions from a PYTHIA calculation and a thermal hydrodynamic model.  
A scaled UA1 parameterization for the $\pi$ production is also shown and is in good
agreement with the PYTHIA
calculation.  Note that the PYTHIA calculation for $\pi$ and the UA1 parameterization are only shown
for $p_{T} > 2$ GeV.
For comparison, also shown are the PHENIX data for $\pi^{+}$ from 5\% central collisions and $\pi^{0}$ from
10\% central collisions for comparison. } 
\label{fig_hydro_pythia_cent10_with_pizerodata}
\end{figure}

In Fig.~\ref{fig_hydro_pythia_cent10_with_pizerodata}, 
the resulting ``Doppler shifted'' transverse momentum distributions 
for $\pi$, $D$ and $B$ hadrons are compared to the pQCD event generator PYTHIA  
 for 10\% central
$Au + Au$ collisions at $\sqrt{s} = 130$ AGeV.  The PYTHIA parameters~\cite{PYTHIA_param} 
for charm and bottom are taken from~\cite{PHENIXe}, where the values 
were tuned to lower energy FNAL charm data and ISR single electron data.
The PYTHIA $\pi$ results use a K factor = 3.5, which agrees well with the
UA1 parameterization scaled to $\sqrt{s} = 130$ GeV and scaled for the $\pi/h$ ratio, as
calculated in~\cite{phenix_highpt}.
For the hydrodynamic model calculation, 
we use a fixed temperature $T=128$ MeV~\cite{heinz} and fit to the PHENIX 
$\pi,K,p$ transverse momentum distributions~\cite{phenix_hadrons}
to determine $\beta_{max} = 0.65$ for central collisions.  These values are
compatible with those previously derived~\cite{jane}. 
The $\pi$ hydrodynamic calculation result is normalized to the PHENIX measured 
$dN/dy(\pi^{+}) = 276 \pm 3$~\cite{phenix_hadrons}, and the $D$ and $B$ results are normalized
to the PYTHIA pQCD integrated dN/dy values.  Different boost profiles are also allowed
and should be considered, including full hydrodynamic model calculations.

Note that in the case of pions, there is a substantial 
difference between the unquenched PYTHIA pQCD prediction and the radial 
hydrodynamic flow results. Also shown in Fig.~\ref{fig_hydro_pythia_cent10_with_pizerodata}
are the PHENIX $\pi^{+}$ yields for 5\% central $Au + Au$ collisions~\cite{phenix_hadrons}
 and the PHENIX $\pi^{0}$ yields for 10\% central $Au + Au$ collisions~\cite{phenix_highpt}.
The data are well described by the hydrodynamic calculation up to $p_{T}\approx 2$ GeV, 
but still fall substantially below the PYTHIA and UA1 parameterization at higher
$p_{T}$.
However, remarkably the difference between these two extreme dynamics
is much less for the heavy open charm and bottom hadrons.  The two models
agree quite well up to $p_T \approx$ 3 and 5 GeV for $D$ and $B$ mesons respectively.

\begin{figure}
\includegraphics[height=.5\textheight]{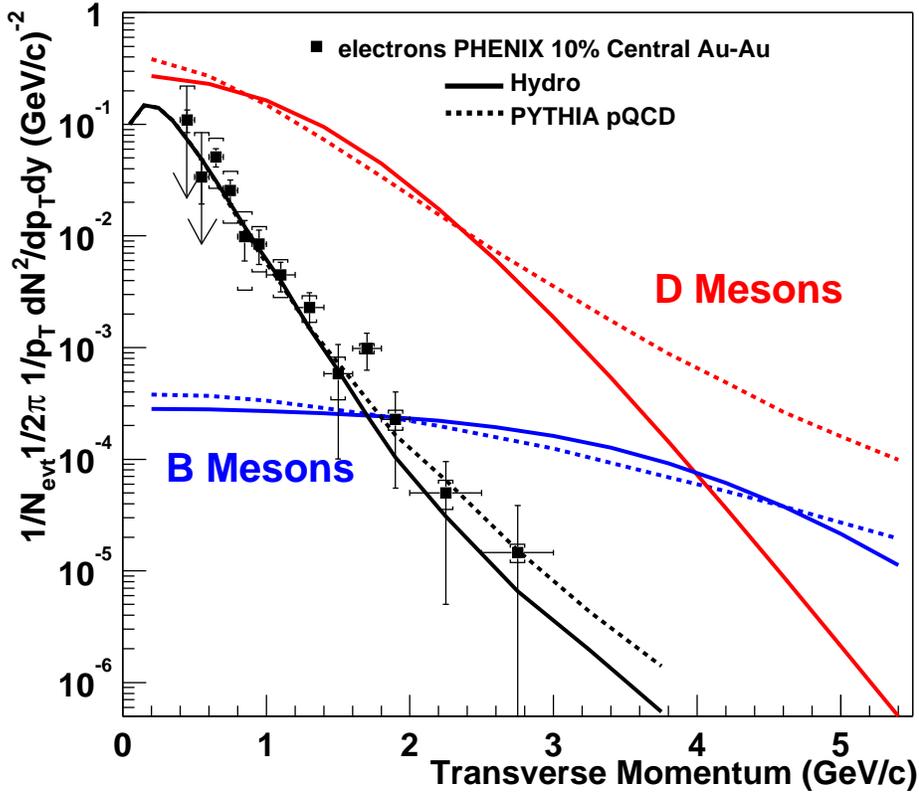}
\caption{The PHENIX 10\% central $Au+Au$ at $\sqrt{s} = 130$ AGeV single electron 
invariant multiplicities as a function of transverse momentum.  
The dashed curves are the PYTHIA calculation 
for $D$ mesons, $B$ mesons and their combined 
resulting decay electrons.  The solid curves are the 
results from the thermal hydrodynamic model.}
\label{fig_elec_hydro_pythia_cent10}
\end{figure}

In Fig.~\ref{fig_elec_hydro_pythia_cent10} we show the hydrodynamic model and  PYTHIA pQCD calculations
for  $D$ mesons, $B$ mesons, and their resulting decay electron distributions for 10\% central
$Au + Au$ collisions at $\sqrt{s} = 130$ AGeV.  
Also shown are the PHENIX measured ``prompt'' single electron distribution~\cite{PHENIXe}.
Both the infinite mean free path pQCD PYTHIA prediction and the zero mean free path
hydrodynamic flow prediction reproduce the electron data.
This is the central observation of this letter.

\begin{figure}
\includegraphics[height=.5\textheight]{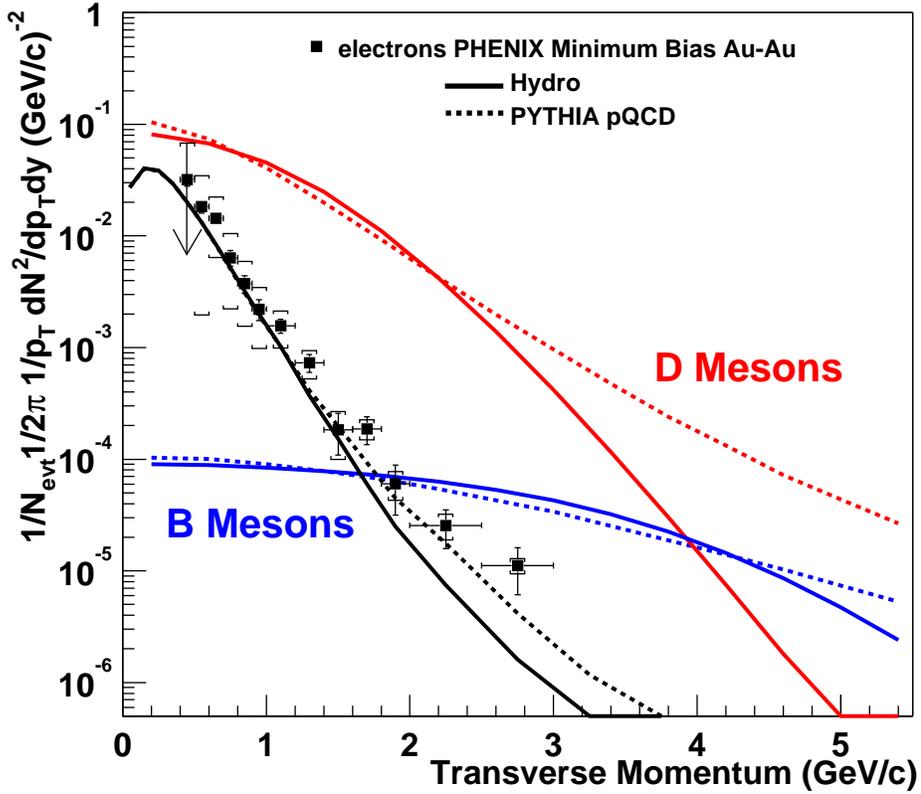}
\caption{The PHENIX ``minimum bias'' (0-92\% central) $Au+Au$  at $\sqrt{s} = 130$ AGeV single electron 
invariant multiplicities as a function of transverse momentum.  
The dashed curves are the PYTHIA calculation 
for $D$ mesons, $B$ mesons and their combined 
resulting decay electrons.  The solid curves are the 
results from the thermal hydrodynamic model.}
\label{fig_elec_hydro_pythia_minbias}
\end{figure}

We have also compared these calculations for ``minimum bias'' (0-92\% central) $Au + Au$ collisions 
at $\sqrt{s} = 130$ AGeV as
shown in Fig.~\ref{fig_elec_hydro_pythia_minbias}.
For the ``minimum bias'' data, PYTHIA pQCD
is scaled by $T_{AA}=6.2~mb^{-1}$ (or equivalently $N_{binary}=246$).  The hydrodynamic calculation
uses temperature (T) and boost ($\beta_{max}$) values for discrete centrality bins and then takes
a weighted average, using the number of binary collisions for each centrality class as the
weight factor.  Assuming a temperature $T=128~MeV$ independent of centrality, we find reasonable 
agreement with the PHENIX $\pi,K,p$ spectra~\cite{phenix_hadrons} with $\beta_{max}$ = 
0.65, 0.65, 0.63, 0.55, 0.25
for centralities 0-5\%, 5-15\%, 15-30\%, 30-60\%, 60-92\%, respectively.  
Again, we find reasonable agreement between hydrodynamics, PYTHIA and the PHENIX experimental data.

It may be a coincidence at RHIC, but the fact is that the initial charm and 
bottom quarks from pQCD may be produced going with the 
flow of light quark and gluons.
This means that even if the heavy quark would  lose 
energy in a {\em static} plasma, they
do not because they are  essentially at rest in the
comoving frame of the flowing plasma. The feedback
of positive energy gain from the comoving matter
makes it impossible for the heavy quark in this momentum range
to change its distribution appreciably. 
The single electron data are thus also consistent 
with the formation of a highly opaque fluid that sweeps 
even the heavy quarks along its strong currents.

If there  exist interactions that 
maintain local thermal equilibrium, could they be strong enough to
bring heavy quarks into chemical equilibrium as well? At lower SPS energies
there is not sufficient time to ``cook up the charm'',
 and chemical equilibrium calculations vastly over-predict the charm yield relative 
to pQCD~\cite{Gorenstein:2000re}. However, at RHIC energies
heavy quark chemistry  may not be as far off
as long as exact charm conservation is taken into 
account in the canonical formulation~\cite{Andronic:2002pj}.
Recent estimates suggest that canonically suppressed charm is in fact
too small by about an order of magnitude relative to pQCD and the PHENIX data~\cite{peterb}.
However, charm and chemistry are exponentially sensitive to
the freeze-out conditions~\cite{Gorenstein:2000re}. 
More work is needed to test this most radical of possibilities.

The single electron data below $p_T \approx 3$ GeV
therefore provides strong motivation to
look to other observables to help differentiate extreme
dynamical scenarios. 
Certainly these models must be confronted with higher statistics single electron
data and a full range of exclusive centrality bins.  This data is expected in the 
near future from the PHENIX experiment in $Au + Au$ at $\sqrt{s} = 200$ AGeV~\cite{ralf_qm}.
Of course more precise data at higher $p_T$ could 
differentiate such  models since pQCD is
only power law suppressed while hydrodynamics is always eventually
exponentially suppressed.  However, this does not rule out a role for
hydrodynamics at the lower $p_T$.


The smoking gun signature for
 hydrodynamic flow would be to observe elliptic flow for $D$ (or $B$)
mesons including of course $J/\psi$.  Unfortunately, the current statistical
method used by PHENIX for measuring charm via single electrons
requires the subtraction of a significant background from Dalitz decay
and conversion electrons.  This ``background'' makes extraction of $v_2$ for heavy
quarks challenging.  Even a displaced vertex tag (from a future
experimental upgrade) would only partially resolve this problem
since the electrons direction is not well correlated
with the $D$ meson direction, thus blurring the orientation for measuring
$v_2$.  It may prove that only a displaced vertex combined
with a complete reconstruction via $D \longrightarrow \pi + K$ will
suffice for such a measurement.  Another possibility is that if 
the plasma is  opaque enough to direct the $J/\psi$ or its 
pre-cursor $c\bar{c}$ into azimuthally asymmetric flow,
then a $v_2$ measurement of $J/\psi$ could serve as the smoking gun.
Note that in this case
the $J/\psi$ may arise  completely from a
late stage ``coalescence''~\cite{Thews:2000rj}.  
This may be the easiest and most direct way 
resolve whether the charm and bottom really flows or are 
simply bystanders that coincidentally  have the same {\em azimuthally
averaged} transverse distribution as the plasma at RHIC.
 
\section{Acknowledgments}  
This work is  supported by the Director, 
Office of Science, Office of High Energy and Nuclear Physics, 
Division of Nuclear Physics, of the U.S. Department of Energy 
under Grant No. DE-FG02-93ER40764 
and DE-FG02-00ER41152 and DE-FG02-86ER40281.  JLN 
acknowledges support from the Alfred P. Sloan Foundation.

\end{document}